\pdfoutput=1
\RequirePackage{lineno}

\documentclass[aps,prl,showpacs,twocolumn,superscriptaddress]{revtex4}

\usepackage{booktabs}
 \usepackage{multirow}
\usepackage{graphicx}    
\usepackage{float}
\usepackage{color}
\usepackage{amsmath}
\usepackage{amssymb}
\usepackage{natbib}
\usepackage{enumitem}
\usepackage{calc}

\begin{document}



\title{Towards a resolution of the proton form factor problem: new
electron and positron scattering data}

\newcommand*{\ODU}{Old Dominion University, Norfolk, Virginia 23529}
\newcommand*{\ODUindex}{27}
\affiliation{\ODU}
\newcommand*{\FIU}{Florida International University, Miami, Florida 33199}
\newcommand*{\FIUindex}{9}
\affiliation{\FIU}
\newcommand*{\ANL}{Argonne National Laboratory, Argonne, Illinois 60439}
\newcommand*{\ANLindex}{1}
\affiliation{\ANL}
\newcommand*{\UTFSM}{Universidad T\'{e}cnica Federico Santa Mar\'{i}a, Casilla 110-V Valpara\'{i}so, Chile}
\newcommand*{\UTFSMindex}{34}
\affiliation{\UTFSM}

\newcommand*{\CANISIUS}{Canisius College, Buffalo, NY}
\newcommand*{\CANISIUSindex}{2}
\affiliation{\CANISIUS}
\newcommand*{\CMU}{Carnegie Mellon University, Pittsburgh, Pennsylvania 15213}
\newcommand*{\CMUindex}{3}
\affiliation{\CMU}
\newcommand*{\CUA}{Catholic University of America, Washington, D.C. 20064}
\newcommand*{\CUAindex}{4}
\affiliation{\CUA}
\newcommand*{\SACLAY}{CEA, Centre de Saclay, Irfu/Service de Physique Nucl\'eaire, 91191 Gif-sur-Yvette, France}
\newcommand*{\SACLAYindex}{5}
\affiliation{\SACLAY}
\newcommand*{\CNU}{Christopher Newport University, Newport News, Virginia 23606}
\newcommand*{\CNUindex}{6}
\affiliation{\CNU}
\newcommand*{\UCONN}{University of Connecticut, Storrs, Connecticut 06269}
\newcommand*{\UCONNindex}{7}
\affiliation{\UCONN}
\newcommand*{\FU}{Fairfield University, Fairfield CT 06824}
\newcommand*{\FUindex}{8}
\affiliation{\FU}
\newcommand*{\FSU}{Florida State University, Tallahassee, Florida 32306}
\newcommand*{\FSUindex}{10}
\affiliation{\FSU}
\newcommand*{\Genova}{Universit$\grave{a}$ di Genova, 16146 Genova, Italy}
\newcommand*{\Genovaindex}{11}
\affiliation{\Genova}
\newcommand*{\GWUI}{The George Washington University, Washington, DC 20052}
\newcommand*{\GWUIindex}{12}
\affiliation{\GWUI}
\newcommand*{\ISU}{Idaho State University, Pocatello, Idaho 83209}
\newcommand*{\ISUindex}{13}
\affiliation{\ISU}
\newcommand*{\INFNFE}{INFN, Sezione di Ferrara, 44100 Ferrara, Italy}
\newcommand*{\INFNFEindex}{14}
\affiliation{\INFNFE}
\newcommand*{\INFNFR}{INFN, Laboratori Nazionali di Frascati, 00044 Frascati, Italy}
\newcommand*{\INFNFRindex}{15}
\affiliation{\INFNFR}
\newcommand*{\INFNGE}{INFN, Sezione di Genova, 16146 Genova, Italy}
\newcommand*{\INFNGEindex}{16}
\affiliation{\INFNGE}
\newcommand*{\INFNRO}{INFN, Sezione di Roma Tor Vergata, 00133 Rome, Italy}
\newcommand*{\INFNROindex}{17}
\affiliation{\INFNRO}
\newcommand*{\INFNTUR}{INFN, sez. di Torino, 10125 Torino, Italy}
\newcommand*{\INFNTURindex}{18}
\affiliation{\INFNTUR}
\newcommand*{\ORSAY}{Institut de Physique Nucl\'eaire, CNRS/IN2P3 and Universit\'e Paris Sud, Orsay, France}
\newcommand*{\ORSAYindex}{19}
\affiliation{\ORSAY}
\newcommand*{\ITEP}{Institute of Theoretical and Experimental Physics, Moscow, 117259, Russia}
\newcommand*{\ITEPindex}{20}
\affiliation{\ITEP}
\newcommand*{\JMU}{James Madison University, Harrisonburg, Virginia 22807}
\newcommand*{\JMUindex}{21}
\affiliation{\JMU}
\newcommand*{\KNU}{Kyungpook National University, Daegu 702-701, Republic of Korea}
\newcommand*{\KNUindex}{22}
\affiliation{\KNU}
\newcommand*{\LPSC}{LPSC, Universit\'e Grenoble-Alpes, CNRS/IN2P3, Grenoble, France}
\newcommand*{\LPSCindex}{23}
\affiliation{\LPSC}
\newcommand*{\UNH}{University of New Hampshire, Durham, New Hampshire 03824-3568}
\newcommand*{\UNHindex}{24}
\affiliation{\UNH}
\newcommand*{\NSU}{Norfolk State University, Norfolk, Virginia 23504}
\newcommand*{\NSUindex}{25}
\affiliation{\NSU}
\newcommand*{\OHIOU}{Ohio University, Athens, Ohio  45701}
\newcommand*{\OHIOUindex}{26}
\affiliation{\OHIOU}

\newcommand*{\URICH}{University of Richmond, Richmond, Virginia 23173}
\newcommand*{\URICHindex}{28}
\affiliation{\URICH}
\newcommand*{\ROMAII}{Universita' di Roma Tor Vergata, 00133 Rome Italy}
\newcommand*{\ROMAIIindex}{29}
\affiliation{\ROMAII}
\newcommand*{\MSU}{Skobeltsyn Institute of Nuclear Physics, Lomonosov Moscow State University, 119234 Moscow, Russia}
\newcommand*{\MSUindex}{30}
\affiliation{\MSU}
\newcommand*{\SCAROLINA}{University of South Carolina, Columbia, South Carolina 29208}
\newcommand*{\SCAROLINAindex}{31}
\affiliation{\SCAROLINA}
\newcommand*{\TEMPLE}{Temple University,  Philadelphia, PA 19122 }
\newcommand*{\TEMPLEindex}{32}
\affiliation{\TEMPLE}
\newcommand*{\JLAB}{Thomas Jefferson National Accelerator Facility, Newport News, Virginia 23606}
\newcommand*{\JLABindex}{33}
\affiliation{\JLAB}

\newcommand*{\EDINBURGH}{Edinburgh University, Edinburgh EH9 3JZ, United Kingdom}
\newcommand*{\EDINBURGHindex}{35}
\affiliation{\EDINBURGH}
\newcommand*{\GLASGOW}{University of Glasgow, Glasgow G12 8QQ, United Kingdom}
\newcommand*{\GLASGOWindex}{36}
\affiliation{\GLASGOW}
\newcommand*{\VT}{Virginia Tech, Blacksburg, Virginia   24061-0435}
\newcommand*{\VTindex}{37}
\affiliation{\VT}
\newcommand*{\VIRGINIA}{University of Virginia, Charlottesville, Virginia 22901}
\newcommand*{\VIRGINIAindex}{38}
\affiliation{\VIRGINIA}
\newcommand*{\WM}{College of William and Mary, Williamsburg, Virginia 23187-8795}
\newcommand*{\WMindex}{39}
\affiliation{\WM}
\newcommand*{\YEREVAN}{Yerevan Physics Institute, 375036 Yerevan, Armenia}
\newcommand*{\YEREVANindex}{40}
\affiliation{\YEREVAN}

\newcommand*{\NOWUK}{University of Kentucky, Lexington, Kentucky 40506}
\newcommand*{\NOWODU}{Old Dominion University, Norfolk, Virginia 23529}
\newcommand*{\NOWINFNGE}{INFN, Sezione di Genova, 16146 Genova, Italy}
\newcommand*{\NOWUF}{University of Florida, Gainesville, FL 32611}
\newcommand*{\NOWJLAB}{Thomas Jefferson National Accelerator Facility, Newport News, VA 23606}


\author{D.~Adikaram}
     \altaffiliation[Current address: ]{\NOWJLAB}
	\affiliation{\ODU}
\author{D.~Rimal}
     \altaffiliation[Current address: ]{\NOWUF}
	\affiliation{\FIU}
\author{L.B.~Weinstein}
	\email[Contact Author \ ]{weinstein@odu.edu}
	\affiliation{\ODU}
\author {B.~Raue} 
	\affiliation{\FIU}
\author {P. Khetarpal} 
	\affiliation{\FIU}
\author{R.P.~Bennett}
	\affiliation{\ODU}	
\author {J.~Arrington}
	\affiliation{\ANL}
\author {W.K.~Brooks}
	\affiliation{\UTFSM}
	
\author {K.P. ~Adhikari} 
\affiliation{\ODU}
\author{A.V.~Afanasev}
\affiliation{GWUI}
% ...
\collaboration{The CLAS Collaboration}
     \noaffiliation

\date{\today}

\begin{abstract}
There is a significant discrepancy between the values of the proton electric form factor, $G_E^p$, extracted using unpolarized and polarized electron scattering. Calculations predict that small two-photon exchange (TPE) contributions can significantly affect the extraction of $G_E^p$ from the unpolarized electron-proton cross sections. We determined the TPE contribution by measuring the ratio of positron-proton to electron-proton elastic scattering cross sections using a  simultaneous, tertiary electron-positron beam incident on a liquid hydrogen target and detecting the scattered particles in the Jefferson Lab CLAS detector. This novel technique allowed us to cover a wide range in virtual photon polarization ($\varepsilon$) and momentum transfer ($Q^2$) simultaneously, as well as to cancel luminosity-related systematic errors.  The cross section ratio increases with decreasing $\varepsilon$ at  $Q^2 = 1.45 \text{ GeV}^2$.  This measurement is consistent with the size of the form factor discrepancy at $Q^2\approx 1.75$ GeV$^2$ and with hadronic calculations including nucleon and $\Delta$ intermediate states, which have been shown to resolve the discrepancy up to $2-3$ GeV$^2$.
\end{abstract}


\maketitle


The electromagnetic form factors describe fundamental aspects of nucleon structure. However, measurements of the ratio of the electric to magnetic proton form factors, $G_E(Q^2)/G_M(Q^2)$,  extracted  using unpolarized and polarized electron elastic scattering data  differ by a factor of three at momentum transfer squared $Q^2\approx 6$ GeV$^2$ \cite{andivahis94,jones00,gayou02,arrington03a,christy04,qattan05,punjabi05,puckett10,puckett11}.  
Until the cause of this surprising discrepancy is  understood, the uncertainty in the form factors can affect the determination of the proton radius, the interpretation of color transparency and $(e,e'p)$ proton knockout measurements, comparisons to isovector and isoscalar nucleon structure calculations from Lattice QCD \cite{Alexandrou06}, and measurements to extract  the flavor-dependent quark contributions to the form factors from parity-violating asymmetries \cite{Beck89}.

One possible explanation for the discrepancy is the presence of two-photon exchange (TPE) effects, where the electron exchanges a virtual photon with the proton, possibly exciting it to a higher state, and then exchanges a second virtual photon, de-exciting the proton back to its ground state. TPE effects  are suppressed by an additional power of the fine structure constant $\alpha = e^2/\hbar \approx 1/137$~\cite{guichon03,blunden03,arrington04d,arrington11b,carlson07}.  Calculations indicate that  TPE effects are small, but increase with electron scattering angle~\cite{blunden05,afanasev05}.  In unpolarized measurements, $G_E$ is extracted from the angular dependence of the elastic cross section at fixed $Q^2$.  For $Q^2>2$ GeV$^2$, the contribution from $G_E$ is less than 10\%, making it very sensitive to even a  small angle-dependent correction. For scattering from a point-like particle, the TPE correction can be calculated exactly \cite{arrington11b}.  However, calculation of the TPE contributions requires a knowledge of {\it all} the baryonic resonance and continuum states that can couple to the two virtual photons.  These corrections  are therefore not yet sufficiently well understood to be applied to the data and are   typically neglected in calculating radiative corrections~\cite{mo69,tsai71,ent01}. 

The most direct way to measure the TPE contributions to the   cross section is by measuring the ratio of positron-proton to electron-proton elastic scattering. However, due to the low luminosity of secondary positron beams,  existing measurements of the $e^+p/e^-p$ cross section ratio are statistically limited  and unable to sufficiently constrain the TPE contribution~\cite{mar68,anderson68,bartel66,arrington04b}. Two new experiments, VEPP-3 at Novosibirsk and OLYMPUS at DESY, will measure the $e^+p$ and $e^-p$  cross sections sequentially using $e^-$ and $e^+$ beams in storage rings~\cite{gramolin12, VEPP-3,Milner2014}. 

This paper describes  a unique technique to compare $e^+p$ and $e^-p$ scattering. Rather than alternating between mono-energetic $e^+$ and $e^-$ beams, we generated a combined electron-positron beam  covering a range of energies and detected  the scattered lepton and struck proton in the CEBAF Large Acceptance Spectrometer (CLAS) at the Thomas Jefferson National Accelerator Facility (Jefferson Lab). This let us simultaneously cover a wide range of momentum transfers and virtual photon polarization,   $\varepsilon = \left[ 1 + 2(1+\tau)\tan^2(\theta/2)\right]^{-1}$,  where $\tau = \frac{Q^2}{4M_p^2}$.  By measuring the $e^+p$ and $e^-p$ elastic cross sections simultaneously, luminosity-related systematic uncertainties cancelled.    

The lepton-proton elastic scattering cross section is proportional to the square of the sum of the Born amplitude and all higher-order QED correction amplitudes. The ratio of $e^\pm p$ elastic scattering cross sections can be written as~\cite{moteabbed13}:
\begin{eqnarray}
R = \frac{\sigma(e^+p)}{\sigma(e^-p)} \approx
\frac{1+\delta_{even}-\delta_{2\gamma}-\delta_{brem}}
     {1+\delta_{even}+\delta_{2\gamma}+\delta_{brem}} \nonumber \\
\approx  1 - 2 ( \delta_{2\gamma} + \delta_{brem})/(1+\delta_{even}) ~,
\label{eq:R}
\end{eqnarray} 
where $\delta_{even}$ is the total charge-even radiative correction factor, and $\delta_{2\gamma}$ and $\delta_{brem}$ are the fractional TPE and lepton-proton bremsstrahlung interference contributions.    After calculating and correcting for the  charge-odd $\delta_{brem}$ term,  the corrected cross section ratio is:
\begin{equation}
\label{eq:R2g}
R' \approx 1 - \frac{2 \delta_{2\gamma}}{(1+\delta_{even})}.
\end{equation}
%


 We produced a simultaneous tertiary  beam of electrons and positrons by using the primary electron beam to produce photons and then using the photons to produce $e^+e^-$ pairs. A $110-140$ nA $5.5$ GeV electron beam struck a $9\times10^{-3} $ radiation length (RL) gold foil to produce a bremsstrahlung photon beam. The electrons were diverted by the Hall-B tagger magnet~\cite{clastagger} into the tagger beam dump. The photon beam  then struck a $9\times10^{-2}$ RL gold foil to produce $e^+e^-$ pairs. The combined photon-lepton beam then entered a three-dipole magnet chicane to horizontally separate the electron, positron and photon beams.  The photon beam was stopped by a tungsten block in the middle of the second dipole.  The lepton beams were recombined into a single beam by the third dipole and then proceeded to a $30$-cm long liquid hydrogen target at the center of CLAS.  For more information on the beam line, see Ref.~\cite{moteabbed13}. The scattered leptons and protons were detected in the CLAS spectrometer~\cite{clasNIM}. 

CLAS is a nearly 4$\pi$ detector. Six superconducting coils  produce an approximately toroidal magnetic field in the azimuthal direction around the beam axis. The sectors between the six magnet cryostats are instrumented with identical detector packages. We used the three regions of drift chambers (DC) ~\cite{clasDCNIM} to measure charged particle  trajectories,  scintillation counters (SC)~\cite{clasSCNIM} to measure time-of-flight (TOF) and forward ($\theta < 45^\circ$) electromagnetic calorimeters (EC)~\cite{clasECNIM} to trigger events.  Additionally,  a Sparse Fiber Monitor, located just upstream of the target, was used to monitor the lepton beam position and stability. A remotely insertable TPE calorimeter (TPECal) located downstream of  CLAS  measured the energy distributions of the individual lepton beams at lower intensity before and after each chicane field reversal. A compact mini-torus magnet placed close to the target  shielded the DC from M{\o}ller electrons.  The CLAS event trigger required at least minimum ionizing energy deposited in the EC in any sector and a hit in the SC in the opposite sector. 

In order to reduce the systematic uncertainties due to potential detector acceptance and incident beam differences, the torus magnet and beam chicane magnet currents were periodically reversed during the run period. The final data set was grouped into four magnet cycles and each magnet cycle contained all possible configurations ($c+t+, c+t-, c-t+, c-t-$ where $c$ and $t$ are the chicane and torus magnet polarities, respectively).  

The symmetric production
of $e^+/e^-$ pairs gives confidence that reversing the  chicane magnet polarity
ensures that the `left beam' luminosity for particles passing on the left side of the chicane is the
same for positive-chicane positrons as for negative-chicane electrons.
This in turn allows us to use the powerful `ratio of ratios' technique.

The ratio between the number of $e^+p$ and $e^-p$ elastic scattering events is calculated in three steps. First, the single ratios are calculated for each magnet configuration as $R_1^{c\pm t\pm} = \frac{N_{e^+p}^{c\pm t\pm}}{N_{e^-p}^{c\pm t\pm}}$. Here $N_{e^\pm p}^{c\pm t\pm}$ are the number of detected elastic events for the different chicane ($c$) and torus ($t$) polarities. The proton detection acceptance and efficiency effects cancel in the single ratio. Next, the double ratios are calculated for each chicane polarity as $R_2^{c\pm } = \sqrt{R_1^{c\pm t+}R_1^{c\pm t-}}$. Any differences in proton and lepton acceptances cancel out in the double ratio. Last, the quadruple ratio is calculated as $R = \sqrt{R_2^{c+}R_2^{c-}}$. The differences in the incident $e^-$ and $e^+$ beam luminosities cancel out in the quadruple ratio \cite{moteabbed13}.   The remaining effects due to lepton-proton correlations and due to the non-reversed magnetic field of the mini-torus were simulated and corrected for as described below.

We applied a series of corrections and cuts to the experimental data to select the elastic $e^\pm p$ events. The systematic deviations in the reconstructed momenta and angles  were studied and corrected. Fiducial cuts in angle and momentum were used to select the region of CLAS with uniform acceptance for both lepton polarities, thus matching the acceptances for $e^+$ and $e^-$.  Contamination from target entrance and exit windows was removed by a 28-cm target vertex cut on both leptons and protons. 

\begin{figure}[htb] 
\begin{center}    
\includegraphics[width=0.49\textwidth,clip=true]{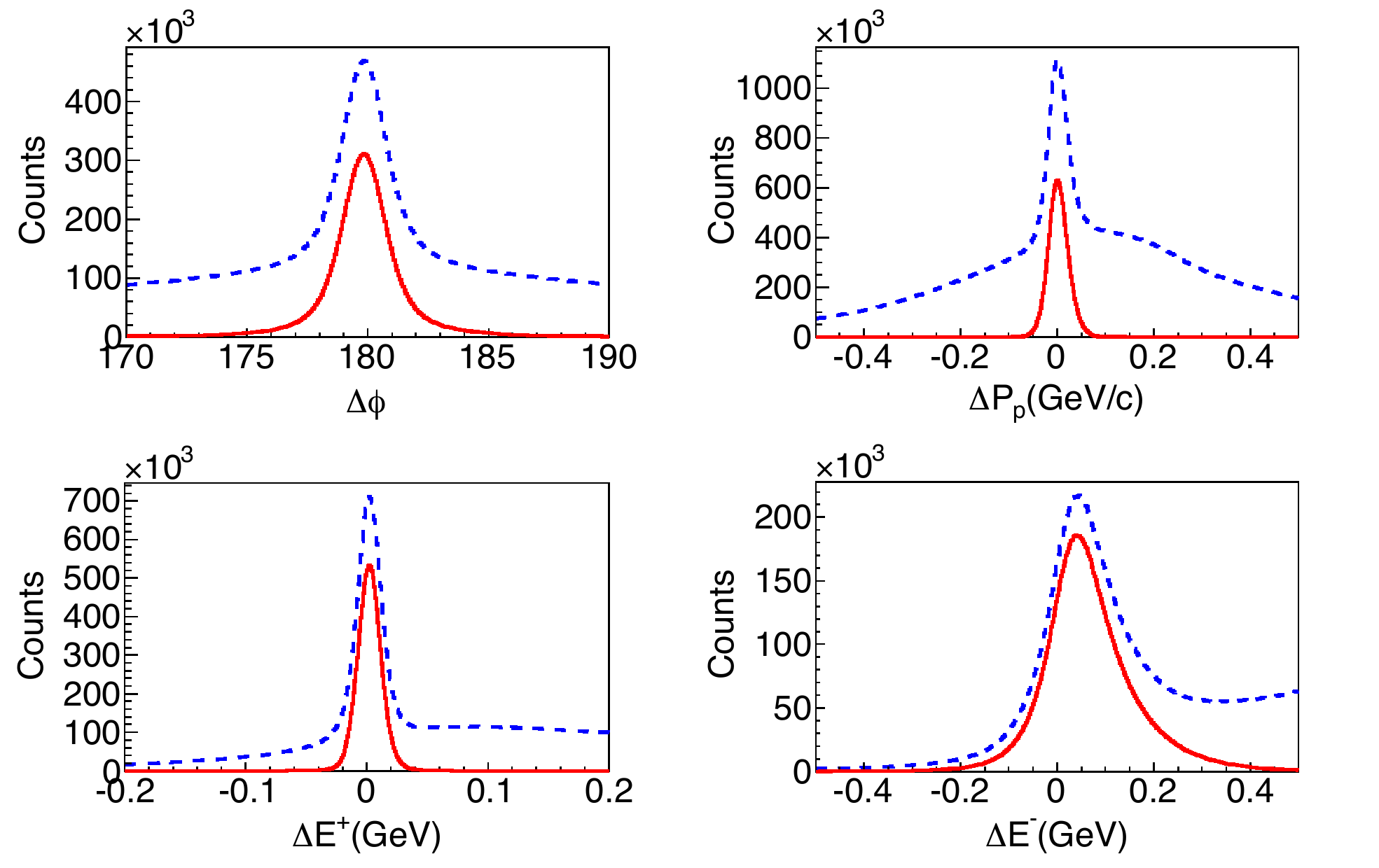}
\caption{(Color online) Number of events as a function of the four variables, $\Delta\phi$, $\Delta P_p$ and $\Delta E^\pm$, before (blue dashed) and after (red) applying the other three elastic cuts on each and summed over all kinematics.}.
\label{fig:KinematicVariables}
\end{center}    
\end{figure}

We calculated the incident lepton energy from the measured scattering angles assuming elastic scattering as $E_l = M_p(\cot(\theta_l/2)\cot\theta_p - 1)$.
Since elastic lepton-proton scattering is kinematically overdetermined when both particles are detected, we applied kinematic cuts on four quantities to select elastic events:  the azimuthal angle difference between the lepton and proton ($\Delta\phi$),  the difference between the incident lepton energy ($\Delta E_l$)  calculated in two different ways,  the difference between the measured and the calculated scattered lepton energy ($\Delta E_l'$) and  the difference between the measured and the calculated recoiling proton momentum ($\Delta p_p$):
\begin{eqnarray}
\Delta\phi &=& \phi_l - \phi_p\nonumber\\
\Delta E_l &=& E_l - (p_l\cos\theta_l + p_p\cos\theta_p)\nonumber\\
\Delta E_l' &=&  \frac{M_pE_l}{E_l(1-\cos\theta_l) + M_p} - E_l' \nonumber\\
\Delta p_p &=& \frac{p_l\sin\theta_l}{\sin\theta_p}  - p_p, \nonumber
\end{eqnarray}
where ($p_l$, $\theta_l$, $\phi_l$) and ($p_p$, $\theta_p$, $\phi_p$) are the measured momenta, scattering angles and  azimuthal angles of the lepton and proton, respectively. The measured scattered lepton energy is $E_l'=p_l$. $\Delta E_l$ and $\Delta E_l'$ are strongly correlated so we applied cuts to $\Delta E^\pm = \Delta E_l \pm \Delta E_l'$. We identified $e^+$ and $p$ kinematically. When this was ambiguous ({i.e.,} when an event with two positive particles passed all four kinematic cuts as either $e^+p$ or $pe^+$) then TOF information was used to identify the $e^+$ and $p$.  We applied $\pm3\sigma$ $Q^2$- and $\varepsilon$-dependent kinematic cuts to select elastic scattering events.  The resulting spectra are remarkably clean (see Fig. \ref{fig:KinematicVariables}).  
	
There is a remnant background seen under the signal, primarily at low $\varepsilon$ and high $Q^2$, even after all other cuts. Since this background is symmetric in $\Delta\phi$, it was estimated by fitting a Gaussian to the tails of the $\Delta\phi$ distribution. We validated the Gaussian shape of the background by comparing it to the background shape determined by the events in the tails of the $\Delta E^-$ distribution. The background was subtracted from the signal before constructing the final cross section ratio. 


 The incident lepton energy distribution rises rapidly from about 0.5 GeV to a peak at about 0.85 GeV and then decreases. We required  $E_{incident}\ge 0.85$ GeV to avoid the region where the  distribution is changing rapidly.  The distributions were slightly different in shape and magnitude ($\approx 10\%$) for different beam chicane polarities, indicating that the chicane was not quite symmetric.  This result is consistent with the incident lepton energy distributions as measured by the TPECal. The TPECal  data  showed that the  $e^+$ energy distribution for positive chicane polarity was identical to the $e^-$ energy distribution for negative chicane polarity (and vice versa).  Therefore differences in $e^+$ and $e^-$  beam luminosities  cancel in the final ratio.  

We matched the detector acceptances by selecting the region of the detector that had a uniform acceptance for both $e^+$ and $e^-$ (fiducial cuts) and by eliminating events that hit a dead channel or would have hit a dead channel if the lepton charge were reversed.  To account for the  non-reversed magnetic field of the mini-torus, we simulated events using GSIM, the CLAS GEANT-based Monte Carlo program. The resulting acceptance correction factors  are all within 0.5$\%$ of unity and were applied to the measured cross section ratios.

\begin{figure}[htb] 
\begin{center}    
\includegraphics[width=0.4\textwidth,clip=true]{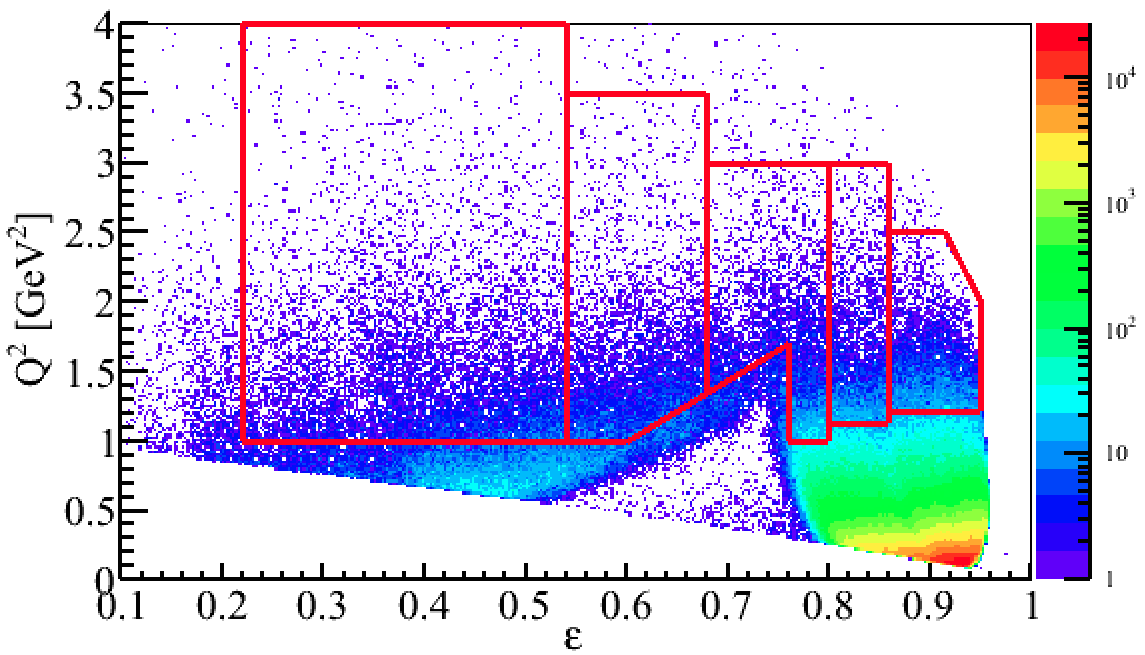}
\caption{(Color online) The number of $e^+p$ elastic scattering events plotted versus $Q^2$ and $\varepsilon$ for positive torus polarity.  The red lines indicate the bin boundaries for the $Q^2\approx 1.45$ GeV$^2$ data.  The hole at $\varepsilon\approx 0.7$ is due to the trigger requirement that at least one of the two particles hit the EC.  The holes for other configurations (negative torus polarity or $e^-p$ events) are smaller.}
\label{fig:Coverage}
\end{center}    
\end{figure}

Our TPE data covered a wide $Q^2$-$\varepsilon$ range (see Fig. \ref{fig:Coverage}).    Small scattering angles $\theta$ correspond to virtual photon polarization $\varepsilon\approx 1$ and large scattering angles correspond to small $\varepsilon$.  The $Q^2 > 1$ GeV$^2$ data were binned into five bins in $\varepsilon$ at an average $Q^2 =1.45$ GeV$^2$. Similarly, the $\varepsilon > 0.8$ data were binned into six $Q^2$ bins at an average $\varepsilon = 0.88$.   The cross section ratio $R$ was measured for each bin. It was then divided by a radiative correction factor equal to the ratio of the $e^+p$ and $e^-p$ radiatively corrected  cross sections  calculated in the modified peaking approximation \cite{ent01} and averaged over each bin by  Monte Carlo integration.  The radiative correction ranged from 0.4\% at $Q^2=0.23$ GeV$^2$ and $\varepsilon=0.88$ to a maximum of $3\%$ at $Q^2 =1.45$ GeV$^2$ and $\varepsilon=0.4$.  The uncorrected, $R$, and radiatively corrected, $R'$, $e^+p/e^-p$ cross section ratios are tabulated in the supplemental information.

Systematic uncertainties  were carefully investigated. The uncertainty due to the target vertex cuts is the difference in the cross section ratios, $R$, between 26 cm and 28 cm target cuts.  The uncertainty due to the fiducial cuts is the difference in $R$ between nominal and  tighter fiducial cuts.  The uncertainty due to the elastic event selection is the difference in $R$ between  $3\sigma$ and $3.5\sigma$ kinematic cuts.  Relaxing the elastic event selection cuts from 3$\sigma$ to 3.5$\sigma$ doubled the background.  Thus the kinematic cut uncertainty also includes the background subtraction uncertainty.  We  varied the background fitting region to determine the additional uncertainty associated with  the fitting procedure.  We  used the six-fold symmetry of CLAS to calculate $R$ independently for each kinematic bin for leptons detected in each of the CLAS sectors (for bins and sectors with good overall efficiency).  We compared the variance of the  measurements with the statistically expected variance to determine the  uncertainty  due to detector imperfections (0.35\%).  The variation in $R$ among the beam chicane magnet cycles was included as an  uncertainty (0.3\%).  The uncertainty in the radiative correction was estimated to be 15\% of the correction (point-to-point) plus a correlated uncertainty of 0.3\% for $Q^2=1.45$ GeV$^2$ and 0.15\% for $\varepsilon=0.88$.  The uncertainties are tabulated in the supplemental information.

Figure~\ref{fig:RatioHiq2} shows  the ratio $R'$  at $Q^2 = 1.45$ GeV$^2$ and at $\varepsilon = 0.88$ compared to hadronic  calculations.  Blunden \textit{et al.} \cite{blunden05} calculated the TPE amplitude using only the elastic nucleon intermediate state. Zhou and Yang~\cite{Zhou2014} considered both the nucleon and the $\Delta(1232)$ in the intermediate state. These calculations bring the form factor ratio extracted from Rosenbluth separation measurements into good agreement with the polarization transfer measurements at $Q^2 < 2-3$ GeV$^2$~\cite{arrington11b} with an additional 1--2\% TPE contribution needed to fully resolve the discrepancy at larger $Q^2$~\cite{arrington07c,Zhou2014}. 

Our data points plus the previous $\varepsilon=0$ point \cite{bouquet68} prefer the hadronic TPE calculation Ref.~\cite{blunden05} by $2.5\sigma$ over the no-TPE ($R'=0$) hypothesis.  A calculation of TPE effects on a structureless point proton \cite{arrington11b} is disfavored by  $5\sigma$.

\begin{figure}[htb] 
\begin{center}    
\includegraphics[width=0.49\textwidth,clip=true]{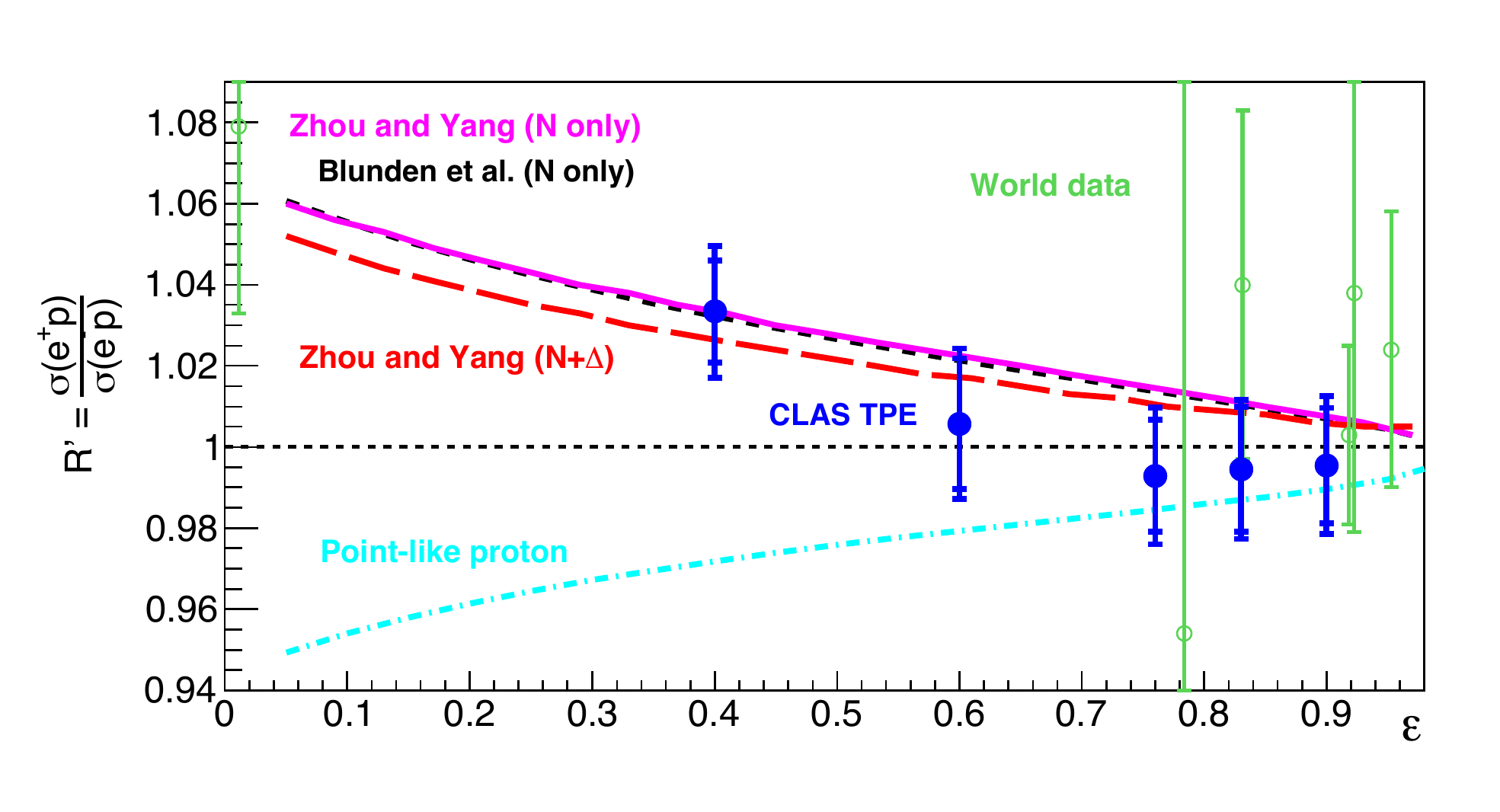}\\
\includegraphics[width=0.49\textwidth,clip=true]{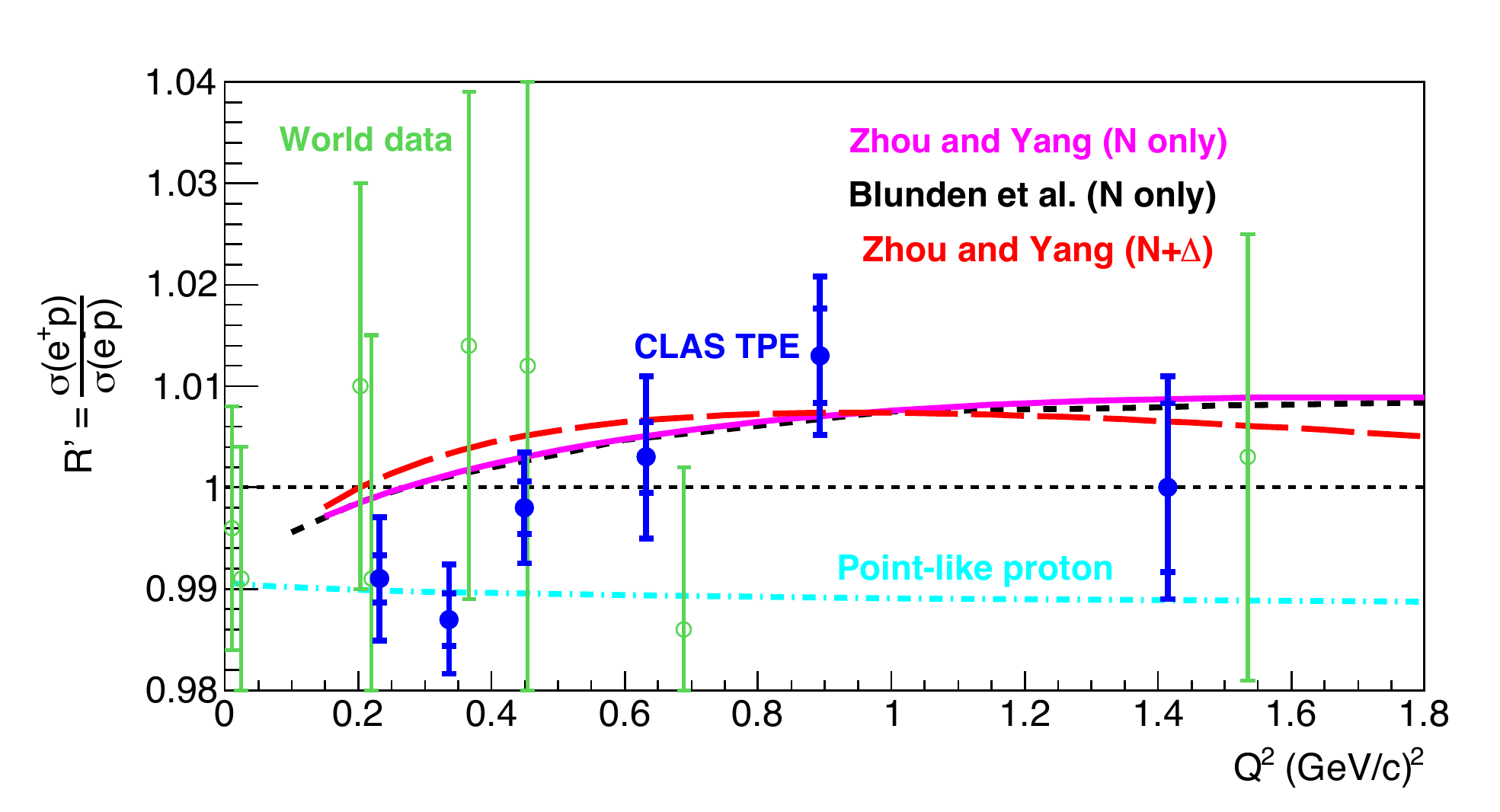}
\caption{(Color online) Ratio of $e^+p$/$e^-p$ cross sections corrected for $\delta_{brem}$ as a function of $\varepsilon$ at $Q^2 = 1.45$ GeV$^2$ (top) and as a function of $Q^2$ at $\varepsilon = 0.88$ (bottom). The filled blue circles show the results of this measurement. The inner error bars are the statistical uncertainties and the outer error bars are the statistical, systematic and radiative-correction uncertainties added in quadrature. The line at $R'=1$ is the limit of no TPE.  The black dotted curve shows the calculation by Blunden \textit{et al.}~\cite{blunden05}. The magenta solid and red dashed curves show the calculation by Zhou and Yang~\cite{Zhou2014} including $N$ only and $N+\Delta$ intermediate states, respectively.  The cyan dot-dashed line shows the calculation of TPE effects on a structureless point proton \cite{arrington11b}.  The open green circles show the previous world data (at $Q^2 > 1$ GeV$^2$ for the top plot)~\cite{arrington04b}.}
\label{fig:RatioHiq2}
\end{center}    
\end{figure}

\begin{figure}[htb] 
\begin{center}    
\includegraphics[width=0.495\textwidth,clip=true]{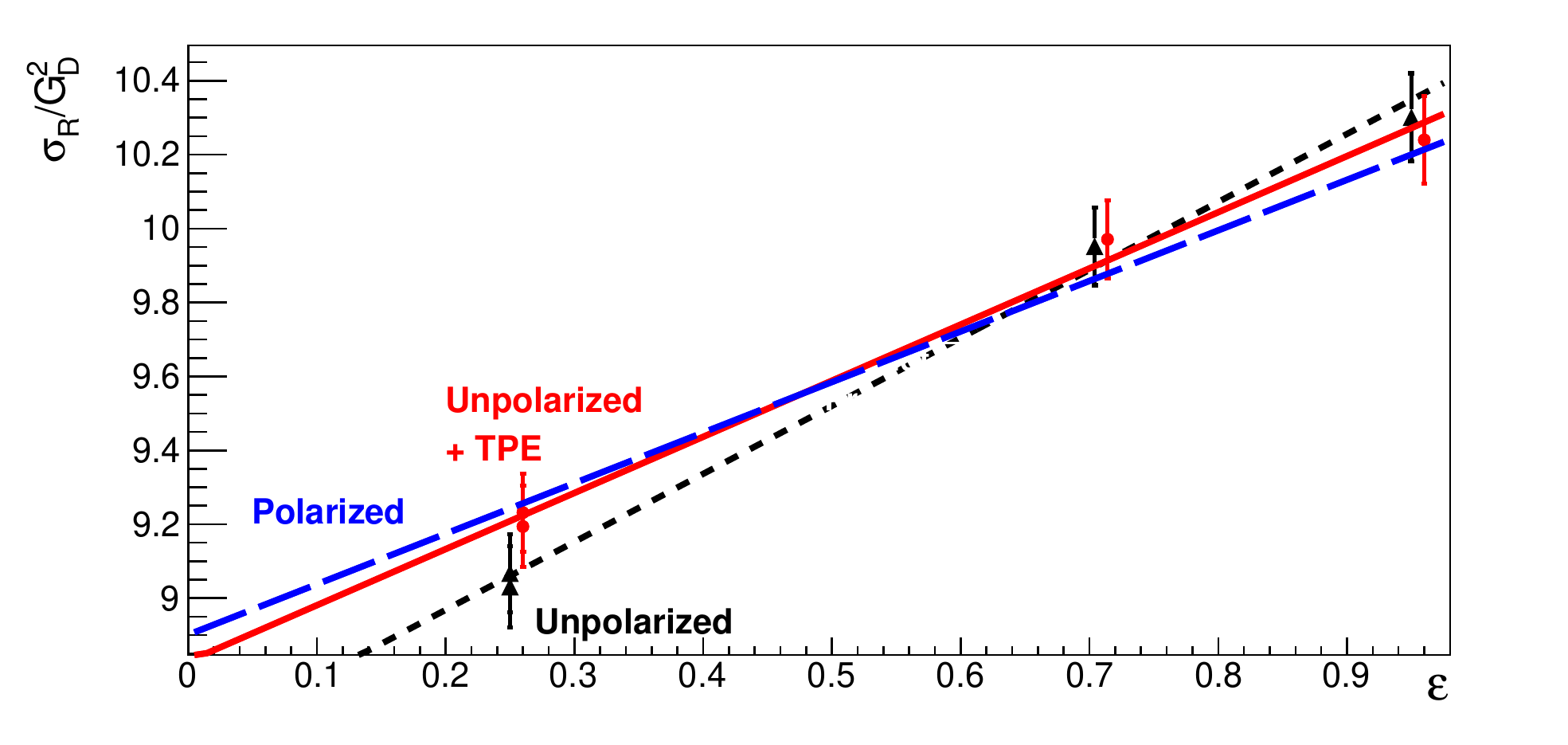}
\caption{(Color online) Reduced cross sections divided by the square of the dipole form factor, $G_D^2= \left(1 + \frac{ Q^2}{0.71 {\rm GeV}^2}\right)^2$, plotted as a function of $\varepsilon$. The black triangles show the original unpolarized measurements from Andivahis \textit{et al.}~\cite{andivahis94} and the red circles show those cross sections corrected by the measured $\delta_{2\gamma}$. The dotted black and solid red lines show the corresponding linear fits where the slope is proportional to $G_E^2$ and the intercept is proportional to $G_M^2$.  The dashed blue line shows the slope expected from the polarized measurement of Punjabi {\it et al.} \cite{punjabi05} (the intercept of this line is arbitrary).}
\label{fig:tpeCorrRosenbluth}
\end{center}    
\end{figure}

We corrected the CLAS TPE cross section ratios at $Q^2=1.45$ GeV$^2$ for the charge-even radiative correction (see Eq. \ref{eq:R2g}) averaged over the appropriate kinematic bins to determine the correction factor $1+\delta_{2\gamma}$.  We fit this to a linear function of $\varepsilon$ and used this to correct  the reduced electron scattering cross sections measured at $Q^2 = 1.75$ GeV$^2$ by Andivahis \textit{et al.}~\cite{andivahis94}:
$	\sigma_R^{corr}(\varepsilon)=\sigma_R(\varepsilon) \left(1+ \delta_{2\gamma}(\varepsilon)\right).$
Fig.~\ref{fig:tpeCorrRosenbluth} shows the original $\sigma_R$ from Andivahis \textit{et al.}~\cite{andivahis94} and the corrected   $\sigma_R^{corr}$ as a function of $\varepsilon$.  The TPE corrections change the proton form factor ratio obtained from the unpolarized data from $\mu_pG_E/G_M = 0.910\pm0.060$ to $0.816\pm0.076$, bringing it into good agreement with the polarized electron scattering result of  $0.789\pm0.042$~\cite{punjabi05}.  This can be seen graphically in Fig.~\ref{fig:tpeCorrRosenbluth} where the slope of the `Unpolarized + TPE' cross section is much closer to that of the polarized results.

In conclusion, we have measured the ratio of $e^+p/e^-p$ elastic scattering cross sections over a wide range of $Q^2$ and $\varepsilon$ using an innovative  simultaneous tertiary $e^+e^-$ beam, detecting the scattered particles in the CLAS spectrometer.  The results are much more precise than previous measurements.  The two photon exchange (TPE)  corrections determined by this experiment from the observed $\varepsilon$-dependence of the $e^+p/e^-p$ cross section ratio at $Q^2=1.45$ GeV$^2$ significantly decrease the proton form factor ratio, $G_E/G_M$, measured by unpolarized elastic scattering data at $Q^2 \sim 1.75$ GeV$^2$ and bring it into good agreement with that determined from  polarized measurements.   Our measurements also support hadronic calculations of two photon exchange (TPE) which resolve the proton form factor discrepancy between polarized and unpolarized electron scattering measurements up to $Q^2 < 2-3$ GeV$^2$~\cite{arrington11b}. 
Verifying the hadronic structure corrections associated with TPE is vital, as such corrections will apply to many other observables \cite{arrington04b,blunden10,arrington07b,tjon09,afanasev05} where direct TPE measurements are not feasible.  

Our results give confidence that the magnetic and electric form-factors of the proton do not scale with one another, implying that there is more involved in the proton's structure than the internal properties of the constituent quarks; for example, angular momentum must reside in orbital motion or in the gluons. 

We  acknowledge the outstanding efforts of the Jefferson Lab staff  (especially  Dave Kashy and the CLAS technical staff)  that made this experiment possible.  This work was supported in part by the U.S. Department of Energy under several grants including DE-FG02-96ER40960 and DE-AC02-06CH11357, the U.S.
 National Science Foundation, the Italian Istituto Nazionale di Fisica
Nucleare, the Chilean Comisi\'on Nacional de Investigaci\'on Cient\'ifica y Tecnol\'ogica 
(CONICYT), the French Centre National de la Recherche Scientifique and Commissariat \`{a} l'Energie Atomique,  the UK Science and Technology Facilities Council (STFC), and the National Research Foundation of Korea.  Jefferson Science Associates, LLC, operates the Thomas Jefferson National Accelerator
Facility for the United States Department of Energy under contract
DE-AC05-060R23177.

\bibliography{tpe}

\end{document}